\documentclass[12pt,aps,prb,preprint,superscriptaddress]{revtex4}
\usepackage{amsmath}
\usepackage{amsfonts}
\usepackage{amssymb}
\usepackage{graphicx}

\begin{document}

\title{A semiclassical optics derivation of Einstein's rate equations}
\author{Robert H\"oppner}
\email{robert.hoeppner@physik.uni-hamburg.de}
\affiliation{Departament d'\`{O}ptica, Universitat de Val\`{e}ncia, Dr.\,Moliner 50,
46100 Burjassot, Spain}
\affiliation{Zentrum f\"ur Optische Quantentechnologien and Institut f\"ur Laserphysik,
Universit\"at Hamburg, 22761 Hamburg, Germany}
\author{Eugenio Rold\'{a}n}
\email{eugenio.roldan@uv.es}
\affiliation{Departament d'\`{O}ptica, Universitat de Val\`{e}ncia, Dr.\,Moliner 50,
46100 Burjassot, Spain}
\author{Germ\'{a}n J. de Valc\'{a}rcel}
\email{german.valcarcel@uv.es}
\affiliation{Departament d'\`{O}ptica, Universitat de Val\`{e}ncia, Dr.\,Moliner 50,
46100 Burjassot, Spain}

\begin{abstract}
\textit{The following article has been submitted to/accepted by the American Journal of Physics. After it is published, it will be found at http://scitation.aip.org/ajp/}
\bigskip

We provide a semiclassical optics derivation of Einstein's rate equations
(ERE) for a two-level system illuminated by a broadband light field, setting
a limit for their validity that depends on the light spectral properties
(namely on the height and width of its spectrum). Starting from the optical
Bloch equations for individual atoms, the ensemble averaged atomic inversion
is shown to follow ERE under two concurrent hypotheses: (i) the
decorrelation of the inversion at a given time from the field at later
times, and (ii) a Markov approximation owing to the short correlation time of
the light field. The latter is then relaxed, leading to effective Bloch
equations for the ensemble average in which the atomic polarization decay
rate is increased by an amount equal to the width of the light spectrum,
what allows its adiabatic elimination for large enough spectral width.
Finally the use of a phase-diffusion model of light allows us to check all
the results and hypotheses through numerical simulations of the
corresponding stochastic differential equations.

\end{abstract}

\maketitle



\section{Introduction}

Lorentz's equation \cite{Lorentz} and Einstein's optical rate equations \cite%
{Einstein} are well known, very valuable, and widely used heuristic models
for the description of light--matter interaction. Both of them occupy an
important place in the early teaching of quantum optics topics too, each of
the models applying to different situations and allowing the study of
different physical problems (refractive index, laser action). Although both
models were formulated before the advent of quantum mechanics, they can be
justified \textit{a posteriori }within the framework of quantum optics
theory in the appropriate limits.\cite{Milonni,Shore,Loudon,Dodd}

Generically speaking, formal derivations of heuristic models from first
principles are important not only for aesthetic reasons and completeness
arguments, but also for the clarification of their applicability domains. In
this respect the situation of Lorentz's and Einstein's models is quite
different: derivations of the former from the optical Bloch equations are
easily found in textbooks,\cite{Milonni,Shore} but this is not the case for
Einstein's rate equations (ERE for short in the following), what results
most surprising given the paramount importance of Einstein's model. Of
course this does not mean that the connection between the optical Bloch
equations and ERE has not been considered, as several quantum optics
textbooks discuss ERE and provide derivations of Einstein's $A$ and $B$
coefficients for spontaneous and stimulated processes,\cite{nota1} see
Refs.\thinspace 4-6. More general treatments can also be found in some
textbooks such as those of Refs.\thinspace 4,\thinspace 8. But these
treatments do not focus on the derivation of ERE (rather on those of
Einstein's $A$ and $B$ coefficients) and, from our viewpoint, do not put
enough emphasis on didactic aspects. We try to close this gap with the
present article.

Einstein proposed his celebrated optical rate equations under the assumption
of a strongly incoherent radiation, and this is the strict meaning of ERE.
However, similar rate equations apply when the atomic line-width
(alternatively, the decay rate of the induced electric dipole) is much
larger than the decay rates of the atomic levels (see, e.g., Ref.\,3), which
occurs, in particular, in many laser systems and this is the reason why this
approach is followed in most laser textbooks. In some quantum optics
textbooks such rate equations are derived from optical Bloch equations
through the adiabatic elimination of the medium polarization and considering
a fully coherent radiation field---unlike Einstein. These derivations lead
to correct rate equations but it is evident that they apply to situations
(strong atomic incoherence) that are far from the ones where ERE do (strong
radiation incoherence). That difference manifests not in the form of the
equations (both are rate equations involving the populations of the two
atomic levels) but in the expressions of the coefficients appearing therein:
Only in the limit of strong radiation incoherence the coefficient governing
stimulated processes in the rate equations is Einstein's $B$ coefficient.

Our treatment below shows, among other things, that the effect of radiation
incoherence manifests as an increase in the decay rate of the \textit{%
ensemble averaged} atomic dipole. Hence when that effective damping rate is
large as compared to the population inversion decay rate (see below a more
rigorous statement) the adiabatic elimination of the electric dipole is
justified, no matter whether the light incoherence or the atomic incoherence
or both are responsible for that largeness. That procedure leads to optical
rate equations in which the radiation--matter coupling constant depends both
on the field spectral width and on the dipole decay rate, which however do
not play a symmetric role. This explains the difference between an adiabatic
elimination based on a large atomic incoherence and that based on a large
light incoherence: Only in the latter Einstein's $B$ coefficient is
obtained. We also notice that in common textbook derivations of the $B$
coefficient a weak field is assumed,\cite{Loudon,Dodd} a limitation absent
in our derivation below.

The rest of this article is organized as follows. In Section II we present
ERE briefly. In Section III we introduce the optical Bloch equations for a
set of atoms and reduce them to an integro-differential equation for the
ensemble averaged atomic inversion. Then, upon applying a decorrelation
approximation between light and atoms, the population inversion dynamics
gets directly connected to the light spectrum. This decorrelation
approximation, which should hold for incoherent light fields, will accompany
us along all our derivations. From that integro-differential equation, in
Section IV we derive ERE by assuming additionally a Markov approximation,
which is justified in the limit of very broad spectra, much broader than the
atomic linewidth (strong light incoherence). Still adopting the Markov
approximation, in Section V we generalize our previous analysis by
considering that either the light spectrum is broad or the atomic line is
(as compared to the inversion relaxation rate), or both, what allows us
to discuss the combined role of light incoherence and atomic incoherence. In
the first case the usual ERE are retrieved with the correct expression for
Einstein's $B$ coefficient while in the second case one recovers the rate
equations used, e.g., in laser modeling. A different approach is used then
in Section VI, where we remove the Markov approximation and transform the
original integro-differential equation into \textit{effective Bloch equations%
}. Importantly in such equations the effective ensemble-averaged atomic
coherence is shown to display a decay rate which is equal to the sum of its
bare decay rate and the width of the light spectrum. Such effective Bloch
equations are analyzed in different limits, in particular when the adiabatic
elimination of the effective atomic coherence is in order. This leads to an
alternative derivation of ERE, as well as to generalized rate equations, and
allows setting the conditions under which rate equations (ERE in particular)
actually hold. In Section VII we study numerically a particularly simple
case (that of a light field having only phase noise), which enables us to
show the validity of the statistical decorrelation assumption used in all
the previous derivations and to discuss a number of questions. Finally the
main conclusions of the work are given in Section VIII.


\section{Einstein's rate equations}

ERE describe the interaction of a broadband isotropic light field with a
two--state atomic system. Einstein\cite{Einstein} postulated three basic
light--matter interaction processes (stimulated absorption and emission, and
spontaneous emission), and established the rate equations governing the
evolution of the populations of each of the atomic states. Denoting by $N_{i}
$ the population of the lower ($i=1$) and upper ($i=2$) atomic states, and
assuming that $N=N_{1}+N_{2}$ is constant and large enough for individual
absorptions and emissions only produce smooth temporal changes in $N_{i}$,
the time evolution of the populations is given by (see Appendix I for a
generalized form) 
\begin{subequations}
\label{Einstein0}
\begin{align}
\frac{\mathrm{d}N_{2}}{\mathrm{d}t}& =-AN_{2}+BW_{21}\left(
N_{1}-N_{2}\right) , \\
\frac{\mathrm{d}N_{1}}{\mathrm{d}t}& =-\frac{\mathrm{d}N_{2}}{\mathrm{d}t},
\end{align}%
where $W_{21}$ is the spectral energy density of the light field at the
atomic transition Bohr frequency $\omega _{21}$, and $A$ and $B\ $are
Einstein's coefficients for spontaneous emission and stimulated processes,
respectively, which neither depend on the field strength nor on time and
verify $A/B=\hbar \omega _{21}^{3}/\pi ^{2}c^{3}$. In terms of the
normalized population inversion $\bar{n}\equiv \left( N_{2}-N_{1}\right) /N$%
, hence $-1\leq \bar{n}\leq 1$, Eqs.\thinspace (\ref{Einstein0}) have the
simpler looking form 
\end{subequations}
\begin{equation}
\frac{\mathrm{d}\bar{n}}{\mathrm{d}t}=-A\left( \bar{n}+1\right) -2BW_{21}%
\bar{n}.  \label{Einstein}
\end{equation}%
We address the reader to Ref.\thinspace 5 for a particularly didactic
presentation and discussion of ERE.


\section{Bloch equations for an ensemble of two--level atoms}

Consider a generic light field whose electric component $\mathcal{\vec{E}}$
we write as 
\begin{equation}
\mathcal{\vec{E}}\left( \mathbf{r},t\right) =\tfrac{1}{2}\left[ \mathbf{E}
\left( \mathbf{r},t\right) +\mathbf{E}^{\ast }\left( \mathbf{r},t\right) %
\right] ,  \label{E}
\end{equation}
where $\mathbf{E}\left( \mathbf{r},t\right) $ is the field
negative-frequency part (that containing terms oscillating as $e^{-i\omega
t} $, $\omega >0$) interacting with a collection of identical two--level
atoms or molecules ($\left\vert 2\right\rangle $ and $\left\vert
1\right\rangle $ will denote the atoms' excited and fundamental states) with
Bohr frequency $\omega _{21}$ and electric dipole matrix elements $%
\left\langle 2\right\vert \mathbf{\hat{\mu}}\left\vert 1\right\rangle
=\left\langle 1\right\vert \mathbf{\hat{\mu}}\left\vert 2\right\rangle
\equiv \mu \mathbf{z}$, which have been taken to be real vectors, aligned
parallel to the Cartesian z-axis, without loss of generality. Working in the
Dirac picture, and after performing the rotating--wave approximation, the
semiclassical optical Bloch equations \textit{for an\ individual atom},
labeled by $\alpha $ and located at $\mathbf{r}_{\alpha }$, can be written
as \cite{Shore,Milonni,Loudon}   
\begin{subequations}
\label{Bloch}
\begin{align}
\frac{\mathrm{d}n_{\alpha }}{\mathrm{d}t}& =-A\left( n_{\alpha }+1\right)
-i\left( \Omega _{\alpha }^{\ast }\sigma _{\alpha }-\Omega _{\alpha }\sigma
_{\alpha }^{\ast }\right) ,  \label{B1} \\
\frac{\mathrm{d}\sigma _{\alpha }}{\mathrm{d}t}& =-\gamma _{\bot }\sigma
_{\alpha }-\frac{i}{2}\Omega _{\alpha }n_{\alpha },  \label{B2}
\end{align}
where $n_{\alpha }=\rho _{22}^{\left( \alpha \right) }-\rho _{11}^{\left(
\alpha \right) }$ and $\sigma _{\alpha }=\rho _{12}^{\left( \alpha \right)
}\exp \left( i\omega _{21}t\right) $ denote, respectively, the population
inversion and the slowly varying atomic coherence of atom $\alpha $
described by its density matrix $\rho ^{\left( \alpha \right) }$, and 
\end{subequations}
\begin{equation}
\Omega _{\alpha }\left( t\right) =\frac{\mu }{\hbar }E_{\mathrm{z}}\left( 
\mathbf{r}_{\alpha },t\right) e^{i\omega _{21}t},  \label{Rabi}
\end{equation}
is the complex Rabi frequency of the light field at the location of atom $%
\alpha $, with $E_{\mathrm{z}}=\mathbf{z}\cdot \mathbf{E}$.

The effect of spontaneous\ emission has been phenomenologically included
through the damping terms, as standard semiclassical theory cannot describe
this process.\cite{nota1} We assume a decay rate $A$ for the population
inversion which implies that the decay rate of the atomic coherence should
be $\frac{1}{2}A$ as we are assuming that only the upper atomic state is
affected by spontaneous emission.\cite{Milonni} However we shall attribute
to $\sigma_{\alpha }$ a decay rate $\gamma_{\bot }=\frac{1}{2}A+\Gamma^{%
\mathrm{dc}}$ which includes an additional decay rate $\Gamma^{\mathrm{dc}}$
describing the effects of dephasing collisions (those affecting the atomic
coherence but not the population inversion). Here we do not consider
radiative collisions (which affect both $n_{\alpha }$ and $\sigma _{\alpha }$
) in order to keep the problem simpler, but they can be easily included (see
Appendix I for a brief discussion on this).

In order to cast our problem in a way similar to ERE (\ref{Einstein}), which
involves just population inversions, we first eliminate the atomic coherence
from (\ref{Bloch}) by integrating formally Eq.\,(\ref{B2}), 
\begin{equation}
\sigma _{\alpha }\left( t\right) =-\frac{i}{2}\int_{0}^{t}\mathrm{d}
t^{\prime }\Omega _{\alpha }\left( t^{\prime }\right) n_{\alpha }\left(
t^{\prime }\right) e^{-\gamma _{\bot }\left( t-t^{\prime }\right) },
\end{equation}
where a transient term, $\sigma _{\alpha
}\left(0\right)\exp\left(-\gamma_{\bot }t\right) $, has been dropped
(alternatively one can take $\sigma _{\alpha }\left(0\right)=0$ without loss
of generality assuming that the interaction is turned on at that instant).
Plugging this into Eq.\,(\ref{B1}) we get 
\begin{align}
\frac{\mathrm{d}n_{\alpha }}{\mathrm{d}t}& =-A\left( n_{\alpha }+1\right) 
\notag \\
& -\mathrm{Re}\int_{0}^{t}\mathrm{d}t^{\prime }\Omega _{\alpha }\left(
t\right) \Omega _{\alpha }^{\ast }\left( t^{\prime }\right) n_{\alpha
}\left( t^{\prime }\right) e^{-\gamma _{\bot }\left( t-t^{\prime }\right) },
\label{npunt}
\end{align}
which is an integro-differential equation for the evolution of the
population inversion of atom $\alpha $.


\subsection{Ensemble averaging}

Equation (\ref{npunt}) rules the population inversion dynamics \textit{of a
single atom}. As we are interested in the average evolution of the whole
system---the ensemble of atoms---, which is the quantity described by ERE,
we introduce the ensemble averaged inversion 
\begin{equation}
\bar{n}\left( t\right) \equiv \left\langle n_{\alpha }\left( t\right)
\right\rangle ,
\end{equation}
where averages are defined as 
\begin{equation}
\left\langle f\left( \mathbf{r}\right) \right\rangle =\frac{1}{N}
\sum_{\alpha =1}^{N}f\left( \mathbf{r}_{\alpha }\right) ,  \label{aver}
\end{equation}
and compute its evolution equation from Eq.\,(\ref{npunt}) as   
\begin{subequations}
\label{Bloch2}
\begin{align}
\frac{\mathrm{d}\bar{n}}{\mathrm{d}t}& =-A\left( \bar{n}+1\right)
-\int_{0}^{t}\mathrm{d}t^{\prime }K\left( t,t^{\prime }\right) e^{-\gamma
_{\bot }\left( t-t^{\prime }\right) },  \label{npunt2} \\
K\left( t,t^{\prime }\right) & =\mathrm{Re}\left\langle \Omega _{\alpha
}\left( t\right) \Omega _{\alpha }^{\ast }\left( t^{\prime }\right)
n_{\alpha }\left( t^{\prime }\right) \right\rangle .  \label{corrn}
\end{align}

Equations (\ref{Bloch2}) describe the average dynamics of the system in an
exact way, in the sense that no approximation has been done on the original
Bloch equations in order to arrive at them. Clearly it is necessary to
evaluate the correlation function $K\left( t,t^{\prime }\right) $ in order
to perform the time integral in Eq.\,(\ref{npunt2}) and then arrive at a
connection of Bloch equations and ERE.


\subsection{The decorrelation approximation}

A direct comparison between Eqs.\,(\ref{Einstein}) and (\ref{Bloch2})
reveals a number of important differences. A main one is that Eq.\,(\ref{Einstein}) has no memory (the time derivative of $\bar{n}$ at time $t$ just
depends on the value of $\bar{n}$ at the same time $t$) while in Eq.\,(\ref{npunt2}) memory effects are present. We will consider this point later
because a previous issue is that in (\ref{npunt2}) it is not only $\bar{n}$
(the ensemble averaged inversion) that rules its evolution but the
individual atomic inversions, $n_{\alpha }$, through the compound
correlation kernel $K.$ A first necessary condition for (\ref{Bloch2}) to
merge with ERE is then that it should be possible to decorrelate $K\left(
t,t^{\prime }\right) $ in (\ref{corrn}) as 
\end{subequations}
\begin{equation}
K\left( t,t^{\prime }\right) \approx \mathrm{Re}\left\langle \Omega _{\alpha
}\left( t\right) \Omega _{\alpha }^{\ast }\left( t^{\prime }\right)
\right\rangle \left\langle n_{\alpha }\left( t^{\prime }\right)
\right\rangle \equiv C\left( t,t^{\prime }\right) \bar{n}\left( t^{\prime
}\right) ,  \label{decorr}
\end{equation}
with 
\begin{equation}
C\left( t,t^{\prime }\right) \equiv \mathrm{Re}\left\langle \Omega _{\alpha
}^{\ast }\left( t\right) \Omega _{\alpha }\left( t^{\prime }\right)
\right\rangle ,  \label{C(t,tp)}
\end{equation}
the field autocorrelation function. Indeed this is a very reasonable
approximation as $K\left( t,t^{\prime }\right) $ contains correlations
between $n_{\alpha }\left( t^{\prime }\right) $ and $\Omega _{\alpha }\left(
t\geq t^{\prime }\right) $, while $\Omega _{\alpha }\left( t\right) $ is
assumed to have a random character. In the following we adopt this
decorrelation approximation (\ref{decorr}), which will be checked numerically
in Section \ref{sec:numerical}.

Using (\ref{decorr}), the Rabi frequency definition (\ref{Rabi}), and
recalling that we are assuming that radiation is isotropic and unpolarized,
which allows writing 
\begin{equation}
\left\langle E_{\mathrm{z}}\left( \mathbf{r},t\right) E_{\mathrm{z}}^{\ast
}\left( \mathbf{r},t^{\prime }\right) \right\rangle =\tfrac{1}{3}%
\left\langle \mathbf{E}\left( \mathbf{r},t\right) \cdot \mathbf{E}^{\ast
}\left( \mathbf{r},t^{\prime }\right) \right\rangle ,  \label{Pol}
\end{equation}Eq.\thinspace (\ref{npunt2}) becomes 
\begin{align}
\frac{\mathrm{d}\bar{n}}{\mathrm{d}t}& =-A\left( \bar{n}+1\right) -\frac{\mu
^{2}}{3\hbar ^{2}}  \notag \\
& \times \mathrm{Re}\int_{0}^{t}\mathrm{d}t^{\prime }\bar{n}\left( t^{\prime
}\right) \left\langle \mathbf{E}\left( \mathbf{r},t\right) \cdot \mathbf{E}%
^{\ast }\left( \mathbf{r},t^{\prime }\right) \right\rangle e^{-\left( \gamma
_{\bot }-i\omega _{21}\right) \left( t-t^{\prime }\right) }.
\end{align}This form is actually very interesting as it allows making direct contact
with the light spectrum via the Wiener-Khintchine theorem as shown in
Appendix II. Making use of Eq.\thinspace (\ref{WK2}) in that Appendix, i.e., 
\begin{equation}
\left\langle \mathbf{E}\left( t\right) \cdot \mathbf{E}^{\ast }\left(
t^{\prime }\right) \right\rangle =\frac{2}{\varepsilon _{0}}\int_{-\infty
}^{+\infty }\mathrm{d}\omega W\left( \omega \right) e^{i\omega \left(
t^{\prime }-t\right) },
\end{equation}we get 
\begin{subequations}
\label{npunt_I}
\begin{align}
\frac{\mathrm{d}\bar{n}}{\mathrm{d}t}& =-A\left( \bar{n}+1\right) -2\frac{%
\pi \mu ^{2}}{3\hbar ^{2}\varepsilon _{0}}\int_{0}^{t}\mathrm{d}t^{\prime }\bar{n}\left( t^{\prime }\right) I\left( t-t^{\prime }\right) e^{-\gamma
_{\bot }\left( t-t^{\prime }\right) },  \label{dn1} \\
I\left( t-t^{\prime }\right) & =\frac{1}{\pi }\mathrm{Re}\int_{-\infty
}^{+\infty }\mathrm{d}\omega W\left( \omega \right) e^{i\left( \omega
_{21}-\omega \right) \left( t-t^{\prime }\right) },  \label{Intensity}
\end{align}%
where a factor $\pi $ has been included in the prefactor of the integral in (\ref{dn1}) for convenience, $W\left( \omega \right) $ denotes the
spectral energy density of the light field and $I\left(t-t^{\prime}\right)$ is  a frequency-shifted field autocorrelation function defined in terms of its spectrum. This is the equation we analyze throughout the rest of this paper.


\section{A first derivation of ERE}

As we already commented in the previous Section ERE have no memory while
Eq.\thinspace (\ref{dn1}) has. But let us put ourselves under the conditions
considered by Einstein: If the light spectrum is broad (a highly incoherent
light field) the field autocorrelation function $I\left( t-t^{\prime }\right) $ defined in (\ref%
{Intensity}) will be a very sharp function around $t^{\prime }=t$. To be
more precise, if we denote by $\Delta $ the width of $W\left( \omega \right) 
$, then $I\left( t-t^{\prime }\right) $ will be effectively zero but for $%
\left\vert t-t^{\prime }\right\vert \lesssim t_{\mathrm{c}}$, where the
coherence time $t_{\mathrm{c}}\sim \Delta ^{-1}$, as follows from the
properties of the Fourier transform. Then, if $\gamma _{\bot }\ll \Delta $
as well one can substitute $\bar{n}\left( t^{\prime }\right) \rightarrow 
\bar{n}\left( t\right) $ and $e^{-\gamma _{\bot }\left( t-t^{\prime }\right)
}\rightarrow 1$ under the integral in (\ref{npunt2}). This constitutes a
Markov approximation indeed as with it one assumes that the lack of
correlation in the field (which implies a large enough amount of randomness
in its evolution) provokes the complete loss of memory of the inversion $%
\bar{n}\left( t\right) $. Under this "light incoherence" dominated scenario
Eqs.\thinspace (\ref{npunt_I}) become, after performing the time
integration, 
\end{subequations}
\begin{equation}
\frac{\mathrm{d}\bar{n}}{\mathrm{d}t}=-A\left( \bar{n}+1\right) -2\frac{\pi
\mu ^{2}}{3\hbar ^{2}\varepsilon _{0}}\bar{n}\int_{-\infty }^{+\infty }%
\mathrm{d}\omega \,W\left( \omega \right) \delta ^{\left( t\right) }\left(
\omega -\omega _{21}\right) ,  \label{dn_delta}
\end{equation}%
where $\delta ^{\left( t\right) }\left( x\right) =\frac{\sin xt}{\pi x}$ is
a Dirac delta like function: It is a peaked function around $x=0$, has a
height equal to $t/\pi $, a width equal to $2\pi /t$, and verifies $%
\int_{-\infty }^{+\infty }\mathrm{d}x\,\delta ^{\left( t\right) }\left(
x\right) =1$. Hence, as soon as $t\gg \Delta ^{-1}$ (which is a very short
time), $W\left( \omega \right) $ can be picked out of the integral as $%
W\left( \omega =\omega _{21}\right) =W_{21}$, leading to 
\begin{subequations}
\label{ERE}
\begin{align}
\frac{\mathrm{d}\bar{n}}{\mathrm{d}t}& =-A\left( \bar{n}+1\right) -2BW_{21}%
\bar{n}, \\
B& =\frac{\pi \mu ^{2}}{3\hbar ^{2}\varepsilon _{0}},  \label{B}
\end{align}%
which coincides with ERE (\ref{Einstein}) and gives the correct result for $B
$ (see Ref.\thinspace 5).


\section{Light incoherence vs atomic incoherence. From ERE to laser rate
equations}

In our previous derivation of ERE we have assumed that the light spectrum
was sufficiently broad as to bring $\bar{n}$ outside the integral in (\ref%
{dn1})---the Markov approximation---and to make the replacement $e^{-\gamma
_{\bot }\left( t-t^{\prime }\right) }\rightarrow 1$. Nevertheless one can
still adopt the Markov approximation without imposing $e^{-\gamma _{\bot
}\left( t-t^{\prime }\right) }\rightarrow 1$, and this is what we face in
this Section. We are hence considering the possibility that, either because
the light spectrum is broad or the atomic line is, or both, (i.e., $\max
\left( \gamma _{\bot },\Delta \right) \gg A$), the function $I\left(
t-t^{\prime }\right) e^{-\gamma _{\bot }\left( t-t^{\prime }\right) }$ under
the integral in (\ref{dn1}) is strongly peaked around $t^{\prime }=t$,
allowing again the substitution $\bar{n}\left( t^{\prime }\right)
\rightarrow \bar{n}\left( t\right) $ under that integral. Hence, under the
Markov approximation Eqs.\thinspace (\ref{npunt_I}) become, after performing
the time integration, 
\end{subequations}
\begin{subequations}
\label{Mark}
\begin{equation}
\frac{\mathrm{d}\bar{n}}{\mathrm{d}t}=-A\left( \bar{n}+1\right) -2\zeta
BW_{21}\bar{n},  \label{EREZ}
\end{equation}%
where $B$ reads as in (\ref{B}), and we defined the dimensionless
coefficient 
\begin{equation}
\zeta =\int_{-\infty }^{+\infty }\frac{\mathrm{d}\omega }{\gamma _{\bot }}%
\frac{W\left( \omega \right) }{W_{21}}\mathcal{L}\left( \frac{\omega -\omega
_{21}}{\gamma _{\bot }}\right) ,  \label{zeta-def}
\end{equation}%
where 
\begin{equation}
\mathcal{L}\left( x\right) =\frac{1}{\pi }\frac{1}{1+x^{2}}.  \label{L}
\end{equation}%
This expression for $\mathcal{L}$ neglects transient terms
proportional to $e^{-\gamma _{\bot }t}$ in the final result, which is a good
approximation as soon as $\gamma _{\bot }t\gtrsim 1$. The Lorentzian $%
\mathcal{L}\left( \frac{\omega -\omega _{21}}{\gamma _{\bot }}\right) $ is
centered at $\omega =\omega _{21}$ and has a width (FWHM) equal to $2\gamma
_{\bot }$, hence it represents the shape of the atomic line: Coefficient $%
\zeta $ is given by the convolution of the (normalized) radiation spectrum
with the atomic absorption spectrum and then provides a measure of the
strength of the interaction.

Remarkably the Markov approximation (in conjunction with the decorrelation
hypothesis) leads naturally to a rate equation description of the system
dynamics. Nevertheless this rate equation (\ref{EREZ}) is not ERE (\ref{ERE}%
) in general because of the presence of coefficient $\zeta$: only if $\zeta=1
$ ERE are obtained. Hence in order to know whether ERE describe the
population dynamics or not it suffices to study the behavior of $\zeta$, as
we do in the following subsections.


\subsection{The light incoherence limit}

In the case when the spectrum is very broad as compared to $\gamma_{\bot }$ (%
$\Delta \gg \gamma_{\bot }$: the "light incoherence" limit considered in the
previous Section) the Lorentzian $\mathcal{L}$ in (\ref{zeta-def}) acts as
an effective Dirac delta by selecting, from $W\left( \omega \right) $, just
the portion around $\omega =\omega_{21}$. Then, if the spectrum is a smooth
function of $\omega $ one can substitute $W\left(\omega \right) $ by $W_{21}$
and, after performing the remaining integral in $\omega $, we get $\zeta=1$,
corresponding to ERE (\ref{ERE}), in agreement with our first derivation.


\subsection{The atomic incoherence limit}

In the opposite limit, i.e., when $\gamma _{\bot }\gg \Delta $, the atomic
linewidth is much broader than the light spectrum; hence one should expect
radiation to behave as effectively coherent. In this case it is $W\left(
\omega \right) $ that acts as an effective Dirac delta in (\ref{zeta-def})
by selecting, from $\mathcal{L}\left( \frac{\omega -\omega _{21}}{\gamma
_{\bot }}\right) $, the portion around the peak of $W$. By analogy to the
previous case we call this the "atomic incoherence" limit. Denoting by $%
\omega _{0}$ the frequency at the peak of $W$, coefficient $\zeta $ takes
the following expression 
\end{subequations}
\begin{equation}
\zeta =\frac{u}{\pi \gamma _{\bot }W_{21}}\mathcal{L}\left( \frac{\omega
_{0}-\omega _{21}}{\gamma _{\bot }}\right) ,
\end{equation}%
where $u=\int_{-\infty }^{+\infty }\mathrm{d}\omega \,W\left( \omega \right) 
$ is the average e.m.\ energy density (see Appendix II). Let us concentrate
on the resonant case, $\omega _{0}=\omega _{21}$, for simplicity, in which
case $\zeta =\frac{u}{\pi \gamma _{\bot }W_{21}}$; taking into account that $%
u$ should be proportional to $\frac{1}{2}\Delta W_{21}$ (a triangular
approximation to the integral $u$)$\,$, we conclude that $\zeta \sim \frac{%
\Delta }{2\gamma _{\bot }}$ and then the interaction is weaker in the
"atomic incoherence" limit than in the "light incoherence" limit (the ERE
case) by a factor $\sim \frac{\Delta }{2\gamma _{\bot }}$. Note that this
"atomic incoherence" limit corresponds to the usual case treated in many
laser textbooks, where the polarization decay rate $\gamma _{\bot }$ is
assumed to be large, which allows the adiabatic elimination of the atomic
coherence (see next Section for an in--depth study of this technique).


\subsection{A bridge between both limits}

\label{subsec:bridge}

To conclude this Section we consider a situation where one can study, in a
continuous fashion, the combined role of the light incoherence and the
atomic incoherence. For this a specific form of the spectrum must be chosen.
We consider the usual Lorentzian form 
\begin{equation}
W\left( \omega \right) =W_{21}\frac{\left(\Delta /2\right) ^{2}}{\left(
\Delta /2\right) ^{2}+\left(\omega -\omega_{21}\right)^{2}},  \label{W_Lor}
\end{equation}
where $\Delta $ represents the width (FWHM) of the light spectrum. Note that
we are using a resonant spectrum; detuned cases can be treated along similar
lines. The final result reads 
\begin{equation}
\zeta =\frac{\Delta }{\Delta +2\gamma_{\bot }}.  \label{Zeta}
\end{equation}

Note that $0<\zeta \leq 1$. This expression contains, as special cases, the
light incoherence limit, in which $\zeta=1$ and ERE (\ref{ERE}) are
recovered, and the atomic incoherence limit where $\zeta=\frac{\Delta}{%
2\gamma_{\bot }}$, all this in agreement with our previous discussion. We
see then that there is a continuous transition from the regime where ERE
apply, dominated by light incoherence, and that typical of laser physics,
dominated by atomic incoherence. Let us insist in that in both cases, and in
intermediate cases as well, the system dynamics is of rate equations type,
see (\ref{EREZ}), but only in the light incoherence limit true ERE are
obtained.


\section{A derivation of ERE based on the adiabatic elimination of effective
Bloch equations. Validity limits of ERE}

\label{sec:effectivebloch}

In the previous Sections we have been able to provide a semiclassical optics
derivation of ERE, with the correct expression for the $B$ coefficient,
under the decorrelation and Markov approximations, whenever the latter is
due to a strong light incoherence. In this case both assumptions are closely
related as both are consequences of the randomness of the incoherent
radiation field.

In this Section we give an alternative derivation of ERE by relaxing the
Markov approximation, which will give us relevant information about the role
of the field spectrum (height and width) and of the dipole relaxation on the
population dynamics. Let us then return to the general case represented by
Eq.\,(\ref{npunt_I}), in which only the decorrelation approximation has been
done. In order to obtain some general, analytic result, a choice must be
made for the form of the spectrum and we use again a Lorentzian one. For the
sake of simplicity we shall also assume that its central frequency is
resonant with the atomic transition, although this is inessential and the
derivations that follow are easily generalized to the non-resonant case.
Using then (\ref{W_Lor}) the intensity (\ref{Intensity}) becomes $I\left(
\tau \right) =$ $\frac{1}{2}\Delta W_{21}\exp \left( -\frac{1}{2}\Delta
\left\vert \tau \right\vert \right) $ and Eq.\,(\ref{npunt_I}) can be
written as 
\begin{equation}
\frac{\mathrm{d}\bar{n}}{\mathrm{d}t}=-A\left( \bar{n}+1\right) -2\zeta
BW_{21}\bar{q},  \label{dndt2}
\end{equation}
where 
\begin{align}
\bar{q}\left( t\right) & \equiv \gamma_{\perp}^{\text{eff}} \int_{0}^{t}\mathrm{d}t^{\prime }\bar{
n}\left( t^{\prime }\right) e^{-\gamma_{\perp}^{\text{eff}} \left( t-t^{\prime }\right) },
\label{def_q} \\
\gamma_{\perp}^{\text{eff}} & =\gamma _{\bot }+\frac{\Delta }{2},  \label{gamma}
\end{align}
the $B$ coefficient reads as in (\ref{B}), and $\zeta $ is given in (\ref%
{Zeta}).

Notice that if $\bar{n}$ varies slowly during a time interval $\frac{1}{\gamma_{\perp}^{\text{eff}}}$
then $\bar{n}\left(t^{\prime }\right) $ can be picked out from the integral
in (\ref{def_q}) at $t^{\prime }=t$ (this is the Markov approximation we
made in the previous Section) and, after performing the remaining integral, $%
\bar{q}\left(t\right) =\bar{n}\left(t\right) $ once $\gamma_{\perp}^{\text{eff}} t\gg 1$, leading
to ERE (\ref{ERE}) but with a modified coefficient for the stimulated
processes when $\zeta $ differs from unity as already discussed.

The integro-differential Eq.\,(\ref{dndt2}) can be easily transformed into a
pair of coupled linear differential equations by taking the time derivative
of (\ref{def_q}),   
\begin{subequations}
\label{Bloch_eff}
\begin{align}
\frac{\mathrm{d}\bar{n}}{\mathrm{d}t}& =-A\left( \bar{n}+1\right) -2\zeta
BW_{21}\bar{q}, \\
\frac{\mathrm{d}\bar{q}}{\mathrm{d}t}& =\gamma_{\perp}^{\text{eff}} \left( \bar{n}-\bar{q}\right)
,  \label{dqdt}
\end{align}
which are \textit{effective Bloch equations} with $\bar{q}$ playing the role
of a kind of normalized average medium polarization. Let us recall that this
set of equations is exact (for a Lorentzian spectrum) but for the
decorrelation assumption (\ref{decorr}), which should hold under a wide
variety of conditions. Furthermore notice that for $\Delta \rightarrow 0$,
i.e., for a fully coherent field hence characterized by a constant Rabi
frequency $\Omega_{0}$ (real without loss of generality), Eqs.\,(\ref%
{Bloch_eff}) are equivalent to Eqs.\,(\ref{Bloch}) (for $\Omega _{\alpha
}\left( t\right) =\Omega_{0}$ $\forall \,\alpha $) upon identifying $\bar{q} 
$ with $iA\sigma /\Omega_{0}$ and $W_{21}$ with $\Omega_{0}^{2}/\Delta $, as
it must be.\cite{nota3}

We see that by assuming the decorrelation hypothesis and by taking a
Lorentzian spectrum for the radiation field, the average Bloch equations for
the atom gas can be reduced to a pair of effective Bloch equations in which
the effective medium polarization decays at a rate $\gamma_{\perp}^{\text{eff}}=\gamma_{\bot
}+\Delta /2=\left(A+\Delta \right) /2+\Gamma^{\mathrm{dc}}$, where $\Delta$
is the FWHM of the light spectrum. Within the range of validity of the
assumptions the above amounts to say that the incoherence of the radiation
field is in some way transferred to the average medium polarization,
manifesting as an increase in its decay rate similar to the effect of
non-radiative collisions, which are accounted for by $\Gamma ^{\mathrm{dc}}$%
. Notice however that $\Delta $ and $\Gamma^{\mathrm{dc}}$ do not enter
symmetrically in the equation of evolution of $\bar{n}$, because of
coefficient $\zeta$, see (\ref{Zeta}).


\subsection{Adiabatic elimination of the effective coherence}

When $\gamma_{\perp}^{\text{eff}}$ is large as compared with $A$---which is the situation we are
interested in---(see below for a more rigorous bound) the effective
coherence $\bar{q}$\ can be adiabatically eliminated from the effective
Bloch equations. We first integrate formally Eq.\,(\ref{dqdt}) thus
recovering (\ref{def_q}). Repeatedly integrating (\ref{def_q}) by parts we
easily get 
\end{subequations}
\begin{equation}
\bar{q}\left(t\right)=\left[1-\left(\gamma_{\perp}^{\text{eff}}\right)^{-1}\frac{\mathrm{d}}{\mathrm{d}t}%
+\left(\gamma_{\perp}^{\text{eff}}\right)^{-2}\frac{\mathrm{d}^{2}}{\mathrm{d}t^{2}}-\ldots \right] \bar{n}
\left(t\right) ,  \label{serie}
\end{equation}
and we see that for large enough $\gamma_{\perp}^{\text{eff}} $ we can approximate $\bar{q}%
\left(t\right)\simeq \bar{n}\left(t\right) $ (notice that this is the same
as taking $\mathrm{d}\bar{q}/\mathrm{d}t=0$ in Eqs.\,(\ref{Bloch_eff}): the
usual adiabatic elimination procedure). Using this result in the equation
for $\bar{n}\left( t\right) $ we retrieve (\ref{EREZ}), which was obtained
under the Markov approximation in that Section. We see then that the latter
and the adiabatic elimination of the effective coherence lead to the same
result, as expected, thus completing our second derivation. Notice that, as
already commented, we obtain a modified coefficient for the stimulated
processes as in general $\zeta \neq 1$. We shall come back to this
difference later.

One of the virtues of this procedure is that it provides us with a simple
tool to determine the conditions under which ERE (or rate equations in
general) are correct, i.e., the conditions under which the above adiabatic
elimination holds.

We can estimate how large $\gamma_{\perp}^{\text{eff}} $ must be by comparing the first two terms
in (\ref{serie}). The necessary condition is $\left\vert\mathrm{d}\bar{n}/ 
\mathrm{d}t\right\vert\ll\gamma_{\perp}^{\text{eff}}\left\vert\bar{n}\right\vert$, which using (%
\ref{Bloch_eff}) leads to 
\begin{equation}
\gamma_{\perp}^{\text{eff}} \gg A,\ \ \ BW_{21}\ll \frac{\gamma_{\perp}^{\text{eff}} }{2\zeta }
\end{equation}
for rate equations to be valid. Using $\gamma_{\perp}^{\text{eff}}=\gamma_{\bot}+\Delta
/2=\left(A+\Delta \right) /2+\Gamma^{\mathrm{dc}}$ the above conditions read
\begin{subequations}
\label{RO0}
\begin{align}
\Delta +2\Gamma ^{\mathrm{dc}}& \gg A, \\
BW_{21}& \ll \frac{\left( A+\Delta +2\Gamma ^{\mathrm{dc}}\right) ^{2}}{%
4\Delta }.
\end{align}

The second condition above sets an upper limit on the field energy density
which is not usually stressed. The first condition implies that the
adiabatic elimination is correct independently on which of the two
quantities $\Delta $ or $\Gamma ^{\mathrm{dc}}$ is the larger one: The
condition is that the effective coherence decay rate be large and not
whether this is due to dephasing collisions or to light incoherence, in
agreement with our initial discussions.

However, importantly, the result is not the same for large $\Delta$ as for
large $\Gamma ^{\mathrm{dc}}$: In the limit of large light incoherence $%
\Delta\gg A,2\Gamma ^{\mathrm{dc}}$ one has $\zeta =1$ and Eq.\,(\ref{EREZ})
exactly coincides with ERE, whilst in the limit of large dephasing
collisions rate, $\Gamma ^{\mathrm{dc}}\gg A,\Delta $, one has $\zeta \simeq
\Delta /\left( 2\Gamma ^{\mathrm{dc}}\right) \ll 1$. This difference can be
rephrased in the following way: For large $\Delta $ the coefficient for
stimulated processes is Einstein's $B$, whilst for large $\Gamma^{\mathrm{dc}%
}$ the coefficient is not $B$ but $B^{\mathrm{eff}}=\zeta B=\Delta
/\left(2\Gamma^{\mathrm{dc}}\right) B$ (see endnote\,10). This is the essential
difference: Einstein's $B$ coefficient appears only when the field is
sufficiently incoherent and not when dephasing collisions are large. This is
so because although dephasing collisions and light incoherence enter
symmetrically in the equation for the effective atomic coherence $\bar{q}$,
through the effective damping rate $\gamma$, they don't in the equation for
the inversion precisely because of the form of coefficient $\zeta$.

We note here that in most textbooks rate equations are derived from optical
Bloch equations (\ref{Bloch}) through the adiabatic elimination of the
medium polarization \textit{without} performing any ensemble averaging that
accounts for the light incoherence. In this case a constant Rabi frequency
is usually assumed, corresponding to coherent radiation, and the atomic
dipole relaxation rate $\gamma_{\bot}$ is assumed to be much larger than the
population one because of the existence of frequent dephasing collisions. As
we have seen this is a correct and legitimate way to derive rate equations,
but it is not the right way for deriving ERE: The price paid with this
simplified presentation is an incorrect expression for the $B$ coefficient.


\subsection{Comparison of rate equations and effective Bloch equations
solutions}

Compared to ERE (\ref{ERE}) the system (\ref{Bloch_eff}) has an extra
equation that allows for a richer dynamics. We shall now compare the
solutions of both models in two different time regimes to obtain a better estimate of the necessary conditions for ERE\ be valid than that of
inequality (\ref{RO0}).


\subsubsection{The short time limit}

For short times after the illumination has been switched on, at $t=0$, the
predictions of both models given by ERE (\ref{EREZ}) and the effective Bloch Eqs.\,(\ref{Bloch_eff}) differ. 

Assuming $\bar{n}\left(0\right) =-1$ and $\bar{q}\left(0\right) =0$ we
easily obtain  
\end{subequations}
\begin{subequations}
\begin{align}
\bar{n}_{\mathrm{ERE}}\left( t\right) & \approx -1+2\zeta BW_{21}t+\mathcal{O}\left(
t^{2}\right)  \label{t_shorta} \\
\bar{n}_{\mathrm{Bloch}}\left( t\right) & \approx -1+\frac{1}{2}\zeta
BW_{21}t^{2}.  \label{t_shortb}
\end{align}
This
means that ERE are overlooking the dynamics at the initial times, see
Fig.\,1(b), in a way similar to what happens with the application of Fermi's
golden rule to the photo-ionization problem.\cite{Fermi} A way to cure the
problem consists in taking $\bar{n}\left( t^{\prime}\right) $ out of the
integral in (\ref{def_q}) when considering the strongly incoherent limit
(i.e., large $\gamma_{\perp}^{\text{eff}} $), and retaining the exact value of the integral,
i.e., approximate $\bar{q}\left(t\right) $ by $\left(1-e^{-\gamma_{\perp}^{\text{eff}} t}\right) 
\bar{ n}\left( t\right) $. With this approximation we get from (\ref{dndt2})
the modified ERE, 
\end{subequations}
\begin{equation}
\frac{\mathrm{d}\bar{n}}{\mathrm{d}t}=-A\left( \bar{n}+1\right) -2\zeta
BW_{21}\bar{n}\left( t\right) \left( 1-e^{-\gamma_{\perp}^{\text{eff}} t}\right) ,  \label{EREmod}
\end{equation}
that predicts a short time evolution as in \eqref{t_shortb}, see Fig.\,1(b). We can now see clearly that the linear time dependence predicted
by ERE (\ref{EREZ}) at short times is an artifact as we are using an approximate equation in a region ($t\lesssim\left(\gamma_{\perp}^{\text{eff}}\right)^{-1}$) where it is not supposed to be valid.


\subsubsection{The long time limit}

\label{subsubsec:longtimelimit}

At long times both the rate equations (\ref{EREZ})---ERE (\ref{ERE}) is special case---and the
effective Bloch Eqs.\thinspace (\ref{Bloch_eff}) reach the same steady state, 
\begin{equation}
\bar{n}\left( \infty \right) =-\frac{A}{A+2\zeta BW_{21}}.  \label{ninf}
\end{equation}%
A stability analysis of ERE provides further insight in how this steady state is approached: It is performed by considering a situation in
which $\bar{n}\left( t\right) $ is close to $\bar{n}\left( \infty \right) $
and seeing how the increment $\delta \bar{n}\left( t\right) \equiv \bar{n}%
\left( t\right) -\bar{n}\left( \infty \right) $ evolves according to (\ref{EREZ}). One trivially gets 
\begin{equation}
\frac{\mathrm{d}}{\mathrm{d}t}\delta \bar{n}=\lambda _{\mathrm{ERE}}\delta 
\bar{n},\ \ \ \ \lambda _{\mathrm{ERE}}=-\left( A+2\zeta BW_{21}\right) ,
\end{equation}%
which leads to a monotonous evolution $\delta \bar{n}\propto \exp \left(
\lambda _{\mathrm{ERE}}t\right) $ in which $\delta \bar{n}$ decreases in
time.

The situation is slightly more involved in the effective Bloch equations case (\ref{Bloch_eff}). Like before, we introduce the
two increments $\delta \bar{n}\left( t\right) \equiv \bar{n}\left( t\right) -%
\bar{n}\left( \infty \right) $ and $\delta \bar{q}\left( t\right) \equiv 
\bar{q}\left( t\right) -\bar{q}\left( \infty \right) $ [Notice that $\bar{q}%
\left( \infty \right) =\bar{n}\left( \infty \right) $] and obtain 
\begin{equation}
\frac{\mathrm{d}}{\mathrm{d}t}%
\begin{pmatrix}
\delta \bar{n} \\ 
\delta \bar{q}%
\end{pmatrix}%
=%
\begin{pmatrix}
-A & -2\zeta BW_{21} \\ 
\gamma_{\perp}^{\text{eff}}  & -\gamma_{\perp}^{\text{eff}}
\end{pmatrix}%
\begin{pmatrix}
\delta \bar{n} \\ 
\delta \bar{q}%
\end{pmatrix}%
,
\end{equation}%
which leads to an evolution of the form $\exp \left( \lambda _{\pm }t\right) 
$ with 
\begin{subequations}
\label{lam}
\begin{eqnarray}
\lambda _{\pm } &=&\frac{-\left( \gamma_{\perp}^{\text{eff}}+A\right) \pm \sqrt{\mathcal{R}}}{2}%
,\  \\
\mathcal{R} &=&\left( \gamma_{\perp}^{\text{eff}} -A\right) ^{2}-8\gamma_{\perp}^{\text{eff}} \zeta B{W}_{21}=\frac{%
\left( \Delta +2\Gamma ^{\mathrm{dc}}-A\right) ^{2}}{4}-4\Delta B{W}_{21},
\end{eqnarray}%
where in the last expression we used (\ref{Zeta}). In the strongly
incoherent limit defined by $\gamma_{\perp}^{\text{eff}} \gg A,\,8BW_{21}$ the above eigenvalues
take the simple form $\lambda _{+}\rightarrow \lambda _{\mathrm{ERE}}$, and $%
\lambda _{-}\rightarrow -\gamma_{\perp}^{\text{eff}} $: The large and negative eigenvalue $\lambda _{-}$ is responsible for a fast evolution that equalizes $\delta \bar{q}$ and $\delta \bar{n}$, and from then on the system evolves as governed by ERE.

For smaller incoherence, however, $\lambda _{\pm }$ may become complex, thus signalling Rabi oscillations with angular frequency equal to $\frac{1}{2}\sqrt{-\mathcal{R}}$, which requires $\mathcal{R}<0$, i.e.,
\end{subequations}
\begin{equation}
BW_{21} >\frac{\left( \Delta +2\Gamma ^{\mathrm{dc}}-A\right) ^{2}}{
16\Delta }\underset{\Delta \gg A,2\Gamma ^{\mathrm{\mathrm{dc}}}}{
\longrightarrow }\frac{\Delta }{16}.  \label{RO1}
\end{equation}%
Hence there are relaxation oscillations in the approach to steady state
whenever the light spectral energy density is large enough, in the sense of
Eq.\thinspace (\ref{RO1}), in stark contrast to the ERE prediction. Nonwithstanding, the observability of the oscillations must be examined because these are damped oscillations according to Eq.\thinspace (\ref{lam}): In order that relaxation oscillations are present their frequency should be larger than, say, half their damping rate. It is easy to check that
Eq.\thinspace (\ref{RO1}) already implies that condition, hence we conclude that relaxation oscillations will occur whenever Eq.\thinspace (\ref{RO1}) is fulfilled.

We conclude that there will be no \textit{qualitative} differences
between the predictions of rate equations and effective Bloch equations,
i.e., that rate equations\ are correct, whenever 
\begin{equation}
BW_{21}<\frac{\left( \Delta +2\Gamma ^{\mathrm{dc}}-A\right) ^{2}}{16\Delta }
,  \label{thr}
\end{equation}
which clarifies the meaning of symbol $\ll $ in inequality (\ref{RO0}). Of
course, that these equations actually correspond to ERE also requires the light incoherence limit $\Delta \gg A,2\Gamma^{\mathrm{dc}}$, in which case the condition becomes $BW_{21}<\frac{\Delta }{16}$.


\section{Stochastic simulation of a field with phase noise}

\label{sec:numerical}

In the previous Sections we have derived ERE (and other rate equations) by
making use of two hypotheses: the decorrelation between the field and the
inversion and the Markov approximation. As the decorrelation hypothesis is crucial for all subsequent derivations and although reasonable because both $\Omega_{\alpha}\left( t\right) $ and $\bar{n}\left( t\right) $ are random variables, we verify its validity for didactic purposes.  For this aim we consider a light field having only phase noise and a tunable degree of incoherence, and take $\Gamma^{\mathrm{dc}}=0$ in order to concentrate on the role of light incoherence. We write the
Rabi frequency at the location of atom $\alpha$ (\ref{Rabi}) as $\Omega_{\alpha}\left( t\right) =\Omega_{0}\exp \left[-i\phi_{\alpha}%
\left(t\right)\right]$, with $\Omega_{0}=\left|\Omega\right|$, and the phase follows the evolution equation 
\begin{equation}
\frac{\mathrm{d}\phi _{\alpha }}{\mathrm{d}t}=\sqrt{\Delta }\xi _{\alpha
}\left( t\right) ,  \label{fip}
\end{equation}
being $\Delta $ a diffusion coefficient that measures the degree of
incoherence of the light field and $\xi_{\alpha}\left(t\right) $ a random function without correlation between atoms. A simple case in which this
is verified is when $\left\{\xi_{\alpha }\left(t\right)
\right\}_{\alpha=1}^{N}$ are different realizations of white Gaussian
noise $\xi\left( t\right) $ which has a mean of $\left\langle \xi_{\alpha
}\left(t\right) \right\rangle=0$ and correlation\cite{nota4} 
\begin{equation}
\left\langle \xi _{\alpha }\left( t\right) \xi _{\alpha }\left( t^{\prime
}\right) \right\rangle =\delta \left( t-t^{\prime }\right) .  \label{corr_xi}
\end{equation}
Note that the averaging operator (\ref{aver}) has the same effect as the
stochastic averaging because we are identifying each atom with a single
realization of the problem, and for this reason we keep the same symbol,
namely "$\left\langle{}\right\rangle$", for the averaging in both pictures.
The problem can thus be described by a set of stochastic
differential equations (SDEs), comprising the Bloch Eqs.\,(\ref{Bloch}) and
Eq.\,(\ref{fip}):   
\begin{subequations}
\label{Stochastic}
\begin{align}
\frac{\mathrm{d}n}{\mathrm{d}t}& =-A\left( n+1\right) -i\Omega _{0}\left(
\sigma e^{i\phi }-\sigma ^{\ast }e^{-i\phi }\right) ,  \label{S1} \\
\frac{\mathrm{d}\sigma }{\mathrm{d}t}& =-\tfrac{1}{2}A\sigma -\frac{i}{2}
\Omega _{0}ne^{-i\phi },  \label{S2} \\
\frac{\mathrm{d}\phi }{\mathrm{d}t}& =\sqrt{\Delta }\xi \left( t\right) .
\label{S3}
\end{align}
Note that we have omitted the atomic subscript $\alpha$ as Eqs.\,(\ref%
{Stochastic}) are interpreted as SDEs.

The field model used here represents laser radiation of
finite linewidth.\cite{Loudon,Mandel} Indeed the spectral energy density $W\left( \omega \right) $ of such field is given by, see Appendix \ref{appendix2}, 
\end{subequations}
\begin{equation}
W\left( \omega \right) =\frac{\Omega _{0}^{2}}{B}\mathrm{Re}\int_{-\infty
}^{+\infty }\mathrm{d}\tau \left\langle e^{i\left[ \phi \left( t+\tau
\right) -\phi \left( t\right) \right] }\right\rangle e^{i\left( \omega
_{21}-\omega \right) \tau },
\end{equation}%
where Eqs.\thinspace (\ref{Rabi}), (\ref{Pol}), (\ref{WK1}) and (\ref{B})
have been used. As $\phi $ is a Wiener process, see Eq.\thinspace (\ref{S3}), its stochastic average is $\left\langle \exp \left( i\left[ \phi \left(
t+\tau \right) -\phi \left( t\right) \right] \right) \right\rangle =\exp
\left( -\frac{1}{2}\Delta \left\vert \tau \right\vert \right) $, see
Ref.\thinspace 9, and therewith we get
\begin{equation}
W\left( \omega \right) =\frac{\Omega _{0}^{2}}{\Delta B}\frac{\left( \Delta
/2\right) ^{2}}{\left( \Delta /2\right) ^{2}+\left( \omega -\omega
_{21}\right) ^{2}},
\end{equation}%
which is a Lorentzian spectrum, as (\ref{W_Lor}), centered at $\omega
=\omega _{21}$ and having a width (FWHM) equal to $\Delta $. In this case 
\begin{equation}
BW_{21}=\Omega _{0}^{2}/\Delta .  \label{W21}
\end{equation}%
Clearly $\Delta $ controls the degree of incoherence as anticipated. We note that ERE (\ref{Einstein}) in this
case, according to (\ref{W21}), reads 
\begin{equation}
\frac{\mathrm{d}\bar{n}}{\mathrm{d}t}=-A\left( \bar{n}+1\right) -2\frac{%
\Omega _{0}^{2}}{\Delta }\bar{n}.  \label{EREphase}
\end{equation} 

We integrated numerically Eqs.\thinspace (\ref{Stochastic}) by means of a
fixed--step midpoint rule algorithm.\cite{Kloeden} The Gaussian noise $\xi \left( t\right) $ was generated with the Box--Muller method.\cite{Box-Muller}
Depending on $\Delta$, we averaged our results over a number of  $N=10^{4}\mathrm{-}10^{6}$ realizations---or atoms---as
many as necessary in order to obtain smooth results. 
We used different values of $\Omega_{0}^{2}$ and $\Delta $ and in all cases we fully confirmed the validity of assumption (\ref{decorr}). We also compared the results of the full time evolution as given by the effective Bloch Eqs.\,(\ref{Bloch_eff}) derived in the previous Section and Eqs.\,(\ref{Stochastic}), as shown in Fig.\,1 where the ensemble averaged population inversion $\bar{n}$ is represented as a function of the dimensionless time $At$ for three values of the Rabi frequency. No difference can be found between the two predictions, which again confirms the validity of the decorrelation hypothesis (\ref{decorr}). Hence we arrive at the conclusion that for a field whose incoherence is solely due to phase noise, Eq.\,(\ref{fip}), the effective Bloch model (\ref{Bloch_eff}) is exact, and hence also that ERE (\ref{EREphase}) are exact for large enough $\Delta$, roughly for $\Delta
\gtrsim 4\Omega_{0},A$, see Eq.\,(\ref{thr}).

We also paid attention to the influence of the number of stochastic
trajectories $N$. In Fig.\thinspace 2 we represent the results of the
numerical integration of Eqs.\thinspace (\ref{Stochastic}) for two different
sets of parameters corresponding to cases with and without relaxation
oscillations (see caption) using different values of $N$ (namely $N=1,10,100,
$ and $1000$). The first feature to be noticed, see Fig.\thinspace 2(a), is
that for $N=1$ and $10$ the trajectories exhibit noisy Rabi oscillations
which disappear for larger $N$. In Fig.\thinspace 2(b) the Rabi oscillations
manifest as relaxation oscillations, but a closer look reveals that also in
this case the \textit{individual} Rabi oscillations manifest well beyond the
disappearance of the relaxation oscillations. In other words, the individual
atomic behavior exhibits Rabi oscillation, and it is the averaging that
removes them.

It is remarkable how the smooth averaged trajectories of Fig.\,1 are
approached as $N$ grows, and it is interesting to note that for small $N$
some oscillations are seen that disappear for larger values of $N$. These
plots suggest that $N=1000$ is a large enough number of trajectories (or
atoms) for the effective Bloch model (\ref{Bloch_eff}) or ERE (\ref{EREphase}
), to be valid, given the chosen parameters. This is more clearly seen in Fig.\,3 where we represent the steady state reached by the system after a long enough transient, as well as its uncertainty, as a function of $N$. Notice that for $N>10^{3}$ the uncertainty is almost negligible.


\section{Conclusions}

\label{sec:conclusions}

We have provided a straight derivation of Einstein's rate equations (ERE)
from the semiclassical optical Bloch equation for an ensemble of closed
two--level atoms or molecules. The derivation has been done by assuming the
statistical decorrelation between the inversion and field correlation
function, and a Markov approximation owed to the (assumed) broad light
spectrum. As well connections between ERE and usual laser rate equations
have been considered, showing that leading to ERE which then have been
analyzed in both the limit where light incoherence dominates, resulting in
ERE with the correct Einstein $B$ coefficient, and the limit where atomic
incoherence dominates, leading to laser rate equations. In the second
derivation, the Markov approximation was replaced by the assumption of a
Lorentzian spectrum in the radiation field. This second derivation led to a
set of effective Bloch equations that contain information about the
radiation spectrum whose bandwidth appears as an increase in the effective
coherence decay rate. Then, for large enough spectral width, ERE are derived
by adiabatically eliminating the effective coherence. Through the analysis
and comparison of the solutions of both the ERE and effective Bloch models
we have derived the conditions under which the former are applicable. We
have discussed the subtle difference that exists between an adiabatic
elimination based on large $\Delta $ (large spectral width) that led to ERE
and provided the correct expression for Einstein's $B$ coefficient, and an
adiabatic elimination based on large $\gamma _{\perp }$ (large atomic
coherence decay rate) which leads to correct laser rate equations but does
not provide a correct $B$ coefficient. In other words: Einstein's $B$
coefficient can only be correctly derived for large spectral width. We think
that this is an important issue from the conceptual and pedagogical points
of view. As well, upper bounds on the field strength have been derived which
should be fulfilled in order that a rate equation description is valid.

Finally, we have studied numerically the decorrelation hypothesis and
checked the different predictions in the special case of a field having only
phase noise. We think our derivations will help students in understanding
more clearly how the ERE model can be justified and under which conditions
it can be applied.

We gratefully acknowledge help from Mar\'{\i}a Gracia Ochoa with some
numerical simulations. This work has been supported by the Spanish
Government and the European Union FEDER through Project FIS2008-06024-C03-01.


\section{Appendix I: Inclusion of radiative collisions}

\label{sec:appendix1}

Consider that besides spontaneous emission from the upper to the lower
atomic state we also take into account the existence of atomic collisions
able of forcing atomic transitions (usually referred to as radiative
collisions). If we denote by $\gamma_{ij}$ the rate at which these
collisions transfer population from level $i$ to level $j$ ($\gamma_{ij}$
depending on the temperature and on the collision cross--section of the
atoms forming the gas), we can rewrite Eqs.\,(\ref{Einstein0}) as 
\begin{equation}
\frac{\mathrm{d}N_{2}}{\mathrm{d}t}=-\left( A+\gamma _{21}\right)
N_{2}+\gamma _{12}N_{1}+BW_{21}\left( N_{1}-N_{2}\right) ,
\end{equation}
and $\mathrm{d}N_{1}/\mathrm{d}t=-\mathrm{d}N_{2}/\mathrm{d}t$. Equation (%
\ref{Einstein}) reads now   
\begin{subequations}
\label{aux}
\begin{align}
\frac{\mathrm{d}\bar{n}}{\mathrm{d}t} &= -\gamma_{\parallel}\left(\bar{n} -
n_{\mathrm{eq}} \right) - 2BW_{21}\bar{n}, \\
\gamma_{\parallel} &= A+\gamma_{21}+\gamma_{12}, \\
n_{\mathrm{eq}} &= -1+2\frac{\gamma_{12}}{\gamma_{\parallel}},
\end{align}
where $n_{\mathrm{eq}}$ is the equilibrium inversion in absence of radiation.

By imposing that at thermal equilibrium $W_{21}$ be given by Planck's
formula,\cite{Loudon} it is easy to see that $\gamma_{12}=\gamma_{21}\exp
\left(-\hbar\omega_{21} /k_{\mathrm{B}}T\right) $, with $k_{\mathrm{B}}$
Boltzmann's constant and $T$ the absolute temperature. This relation between 
$\gamma_{21}$ and $\gamma_{12}$ means that, in thermal equilibrium, for each
collision induced atomic excitation there must be a corresponding
deexcitation.

We can also add radiative collisions in the optical Bloch equations in a
similar way. Now we must write \cite{Milonni}   
\end{subequations}
\begin{subequations}
\label{aux1}
\begin{align}
\frac{\mathrm{d}n_{\alpha }}{\mathrm{d}t}& =-\gamma _{||}\left( n_{\alpha
}-n_{eq}\right)-i\left( \Omega _{\alpha }^{\ast }\sigma _{\alpha }-\Omega
_{\alpha }\sigma _{\alpha }^{\ast }\right) , \\
\frac{\mathrm{d}\sigma _{\alpha }}{\mathrm{d}t}& =-\gamma _{\bot }\sigma
_{\alpha }-\frac{i}{2}\Omega _{\alpha }n_{\alpha },
\end{align}
with $\gamma_{\bot }=\gamma_{||}/2+\Gamma^{\mathrm{dc}}$. Notice that, in
general, $\gamma_{12}\simeq 0$ is a good approximation.

Let us insist in that the above Bloch equations, and also Eq.\,(\ref{aux}),
apply to a \textit{closed }two level atomic system. If the system is assumed
to be open (i.e., if relaxation processes can bring the atom into atomic
states different from the two that interact with the light field) the
relaxation terms need to be appropriately generalized (see, e.g., Ref.\,3).

By following exactly the same lines we have followed in this article, the
generalized Einstein Eq.\,(\ref{aux}) is derived from Eqs.\,(\ref{aux1}).


\section{Appendix II: The Wiener-Khintchine theorem and its application to
EM radiation}
\label{appendix2}

The Wiener--Khintchine theorem relates the spectral energy density with the
field's autocorrelation function (see, e.g., Ref.\,5). It can be put in
either form   
\end{subequations}
\begin{subequations}
\label{WK}
\begin{gather}
W\left( \omega \right) =\frac{\varepsilon _{0}}{4\pi }\int_{-\infty
}^{+\infty }\mathrm{d}\tau \left\langle \mathbf{E}\left( t\right) \cdot 
\mathbf{E}^{\ast }\left( t+\tau \right) \right\rangle e^{-i\omega \tau },
\label{WK1} \\
\left\langle \mathbf{E}\left( t\right) \cdot \mathbf{E}^{\ast }\left( t+\tau
\right) \right\rangle =\frac{2}{\varepsilon _{0}}\int_{-\infty }^{+\infty } 
\mathrm{d}\omega W\left( \omega \right) e^{i\omega \tau },  \label{WK2}
\end{gather}
and this Appendix II is devoted to demonstrate this. The derivation bases on
considering the e.m. field defined inside a fictitious volume (a cube of
side $L$) with periodic boundary conditions allowing for the existence of
traveling waves as in free space, and then letting $L\rightarrow \infty $,
which allows passing to the continuum.

The electric and magnetic fields in such case can be written in their more
general form as  
\end{subequations}
\begin{subequations}
\begin{equation}
\mathcal{\vec{X}}\left( \mathbf{r},t\right) =\sum_{\mathbf{n}}\sum_{\sigma } 
\mathcal{\vec{X}}_{\mathbf{n},\sigma }e^{i\left( \mathbf{k}_{\mathbf{n}
}\cdot \mathbf{r}-ck_{\mathbf{n}}t\right) }+c.c.,  \label{defEB}
\end{equation}
where $\mathcal{\vec{X}}=\mathcal{\vec{E}},\mathcal{\vec{B}}$,  
\begin{align}
\mathcal{\vec{E}}_{\mathbf{n},\sigma }& =L^{-3/2}\mathbf{e}_{\sigma }\left( 
\mathbf{n}\right) A_{\mathbf{n},\sigma }, \\
\mathcal{\vec{B}}_{\mathbf{n},\sigma }& =\left( ck_{\mathbf{n}}\right) ^{-1} 
\mathbf{k}_{\mathbf{n}}\times \mathcal{\vec{E}}_{\mathbf{n},\sigma },
\end{align}
$\mathbf{n\in \mathbb{Z}}^{3}$, $\sigma =1,2$, $\mathbf{k}_{\mathbf{n}}= 
\frac{2\pi }{L}\mathbf{n}$, $k_{\mathbf{n}}=$ $\left\vert \mathbf{k}_{ 
\mathbf{n}}\right\vert $, the polarization unit vectors (which we choose to
be real --linear polarization basis-- without loss of generality) verify $%
\mathbf{e}_{\sigma }\left(\mathbf{n}\right) \cdot \mathbf{e}_{\sigma
^{\prime }}\left(\mathbf{n}\right)=\delta_{\sigma ,\sigma ^{\prime }}$, and
we take $\mathbf{e}_{\sigma }\left(-\mathbf{n}\right) =\mathbf{e}_{\sigma
}\left( \mathbf{n}\right) $ by convention. First we compute the average e.m.
energy density contained in the volume, 
\end{subequations}
\begin{equation}
u=\frac{\varepsilon _{0}}{2}\left\langle \mathcal{\vec{E}}^{2}+c^{2}\mathcal{%
\ \vec{B}}^{2}\right\rangle =\frac{1}{L^{3}}\int_{L^{3}}\mathrm{d}^{3}r\frac{
\varepsilon _{0}}{2}\left( \mathcal{\vec{E}}^{2}+c^{2}\mathcal{\vec{B}}
^{2}\right).
\end{equation}
Upon using (\ref{defEB}) and taking into account that  
\begin{subequations}
\begin{align}
\frac{1}{L^{3}}\int_{L^{3}}\mathrm{d}^{3}re^{i\left( \mathbf{k}_{\mathbf{n}}+%
\mathbf{k}_{\mathbf{n}^{\prime }}\right) \cdot \mathbf{r}}& =\delta _{ 
\mathbf{n},-\mathbf{n}^{\prime }}, \\
\left[ \mathbf{k}_{\mathbf{n}}\times \mathbf{e}_{\sigma }\left( \mathbf{n}%
\right) \right] \cdot \left[ \mathbf{k}_{-\mathbf{n}}\times \mathbf{e}%
_{\sigma ^{\prime }}\left( \mathbf{n}\right) \right] & =-k_{\mathbf{n}%
}^{2}\delta _{\sigma ,\sigma ^{\prime }},
\end{align}
one obtains, after little algebra, 
\end{subequations}
\begin{equation}
u=2\varepsilon _{0}\sum_{\mathbf{n}}\sum_{\sigma }L^{-3}\left\vert A_{ 
\mathbf{n},\sigma }\right\vert ^{2}.
\end{equation}
Finally passing to the continuum [$\mathrm{d}^{3}k=\left( 2\pi /L\right)
^{3} $] and working in spherical coordinates [$\mathrm{d}^{3}k=k^{2}\mathrm{d%
}k\mathrm{d}^{2}\Omega $] we get straightforwardly  
\begin{subequations}
\begin{align}
u& =\int_{0}^{+\infty }\mathrm{d}\omega W\left( \omega \right) , \\
W\left( \omega \right) & =\frac{\varepsilon _{0}\omega ^{2}}{4\pi ^{3}c^{3}}
\sum_{\sigma }\int_{4\pi }\mathrm{d}^{2}\Omega \left\vert A_{\sigma }\left( 
\mathbf{k}\right) \right\vert _{k=\omega /c}^{2},  \label{W_calc}
\end{align}
where $\mathrm{d}\omega =c\mathrm{d}k$, and $\left\vert A_{\sigma }\left( 
\mathbf{k}\right) \right\vert^{2}=\left\vert A_{\mathbf{n},\sigma
}\right\vert^{2}$ such that $\mathbf{n}=\mathbf{k/}\left(2\pi /L\right) $.

Next we compute the correlation $\left\langle \mathbf{E}\left( \mathbf{r}
,t\right) \cdot \mathbf{E}^{\ast }\left( \mathbf{r},t+\tau \right)
\right\rangle $, which following the previous lines can be written as 
\end{subequations}
\begin{equation*}
\left\langle \mathbf{E}\left( \mathbf{r},t\right) \cdot \mathbf{E}^{\ast
}\left( \mathbf{r},t+\tau \right) \right\rangle =\frac{1}{2\pi ^{3}c^{3}}
\sum_{\sigma }\int_{4\pi }\mathrm{d}^{2}\Omega \int_{0}^{+\infty }\mathrm{d}
\omega \omega ^{2}\left\vert A_{\sigma }\left( \mathbf{k}\right) \right\vert
_{k=\omega /c}^{2}e^{i\omega \tau }.
\end{equation*}
Taking the Fourier transform of the above expression we get 
\begin{equation*}
\int_{-\infty }^{+\infty }\mathrm{d}\omega ^{\prime }\left\langle \mathbf{E}
\left( \mathbf{r},t\right) \cdot \mathbf{E}^{\ast }\left( \mathbf{r},t+\tau
\right) \right\rangle e^{-i\omega ^{\prime }\tau }=\frac{\omega ^{\prime 2}}{
\pi ^{2}c^{3}}\sum_{\sigma }\int_{4\pi }\mathrm{d}^{2}\Omega \left\vert
A_{\sigma }\left( \mathbf{k}\right) \right\vert _{k=\omega ^{\prime }/c}^{2},
\end{equation*}
which, compared with (\ref{W_calc}), yields (\ref{WK}). $\square $


\section{Figure captions}

\subsection{Figure 1}

Evolution of the ensemble averaged population inversion $\bar{n}$ as a
function of the dimensionless time $At$. In (a) $\Omega_{0}=4$ and we used
the three values of $\Delta$ indicated in the figure. These results have
been obtained by numerically integrating Eqs.\,(\ref{Bloch2}) and the
results coincide exactly with those provided by the effective Bloch Eqs.\,(%
\ref{Bloch_eff}). In (b) we used $\Omega_{0}=\sqrt{11}\;\left(\zeta
BW_{21}=2\right) $ and $\Delta =5$, and we have represented the predictions
of the Bloch model (Eqs.\,(\ref{Bloch2}), blue line), of ERE (Eqs.\,(\ref%
{EREphase}), red line), and of the modified ERE (Eqs.\,(\ref{EREmod}), brown
line). The inset shows the different predictions for short times (see text).

\subsection{Figure 2}

Evolution of the averaged inversion $\bar{n}$ obtained with the Bloch model
Eqs.\,(\ref{Bloch2}) for (a) $\Omega_{0}=2$ and $\Delta=10$, and (b) $%
\Omega_{0}=6$ and $\Delta=1$, for several values of the number of atoms $N$.
In (b) the trajectories for $N=100$ and $N=10000$ (not labeled for the sake
of clarity) are so close each other that we plotted the former with dashed
line in order to distinguish them.

\subsection{Figure 3}

Average population inversion $\pm $ its standard deviation after $N$
trajectories for the steady state using (a) $\Delta=10,\,\Omega_{0}=2$ (no
transient oscillations) and (b) $\Delta =1,\Omega_{0}=6$ (strong transient
oscillations).


\begin{figure}[ht]
\begin{center}
\scalebox{0.80}{ 
		\includegraphics{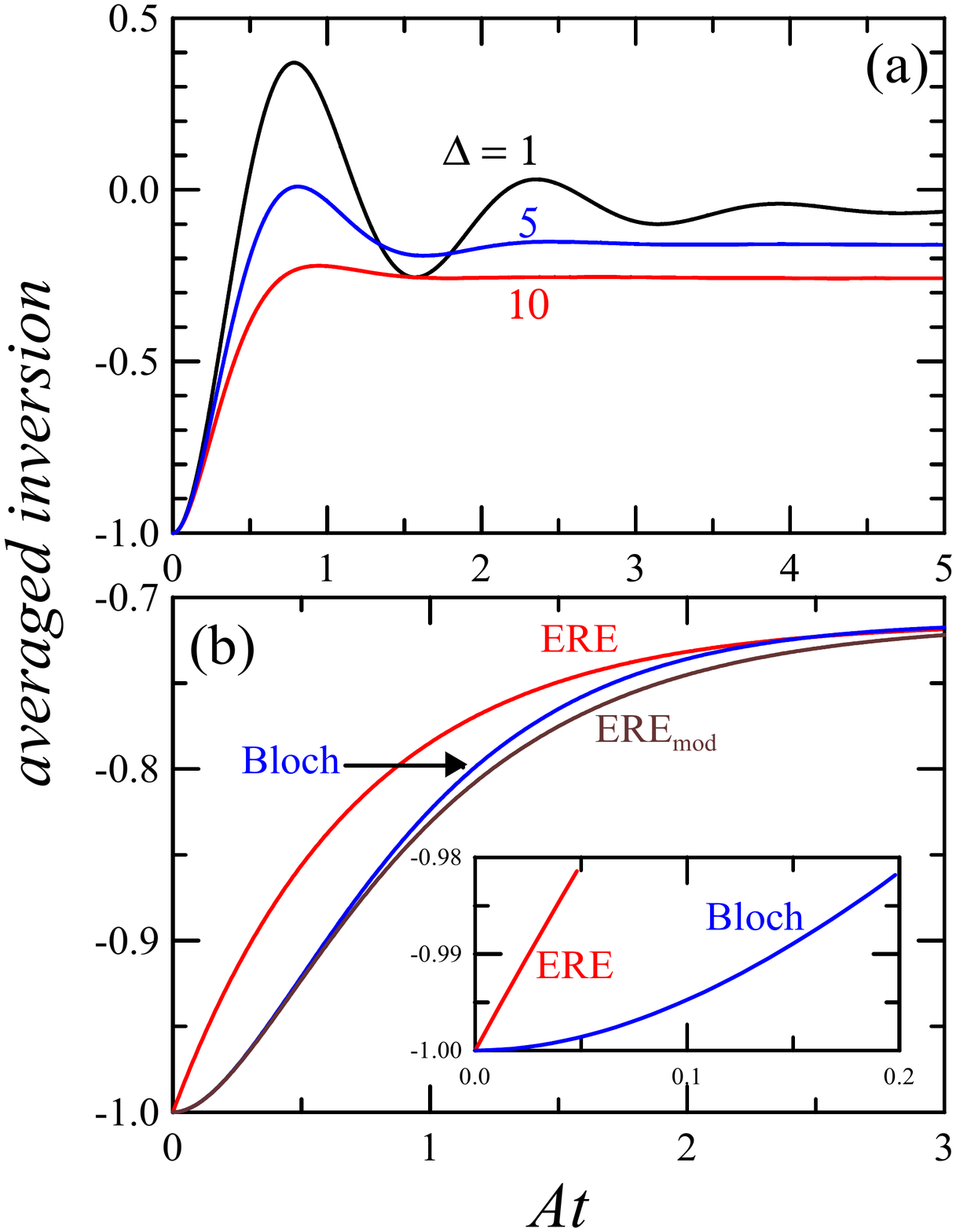}} 
\end{center}
\caption{}
\label{fig:FIG1}
\end{figure}
\begin{figure}[ht]
\begin{center}
\scalebox{0.80}{ 
		\includegraphics{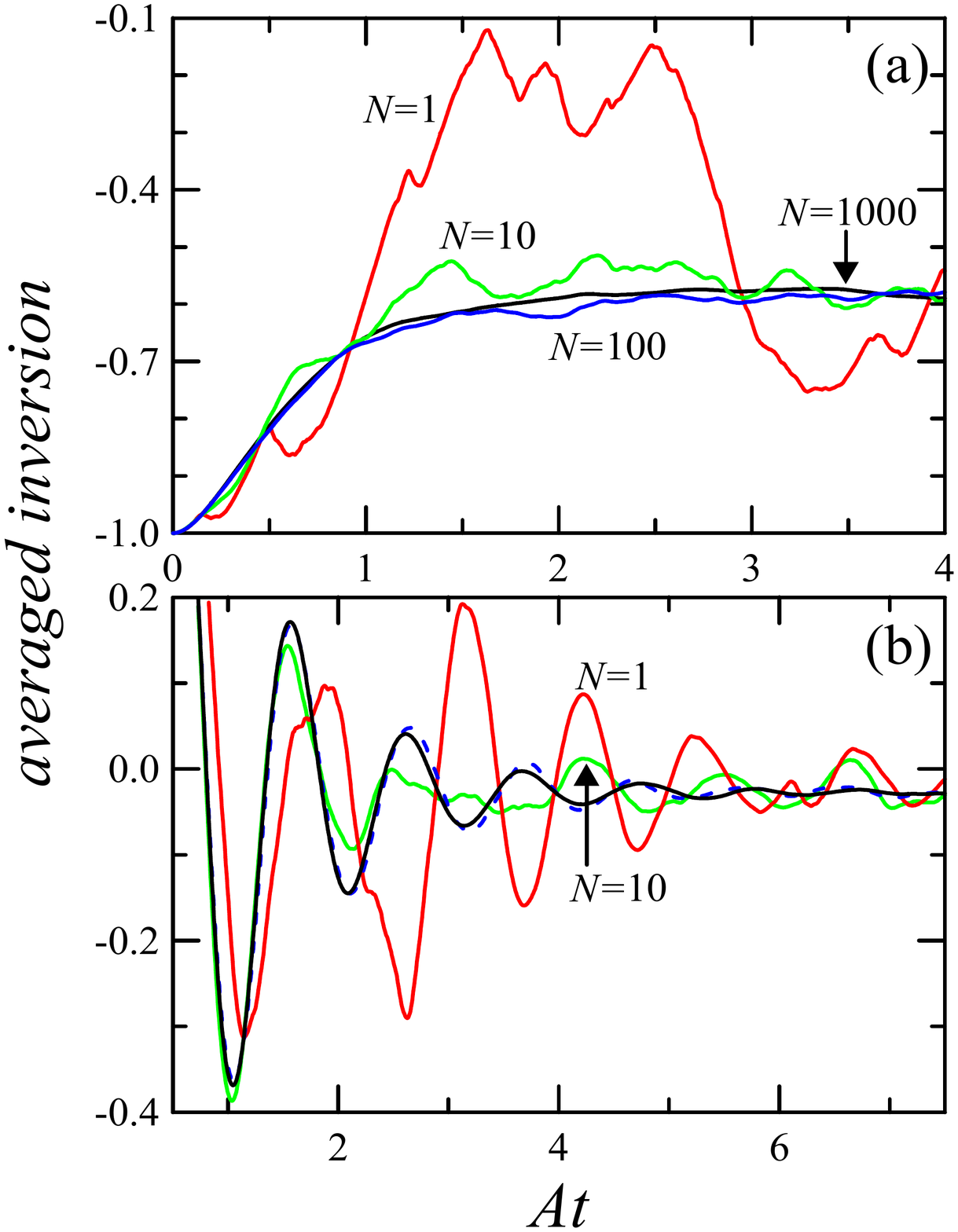}} 
\end{center}
\caption{}
\label{fig:FIG2}
\end{figure}
\begin{figure}[ht]
\begin{center}
\scalebox{0.80}{ 
		\includegraphics{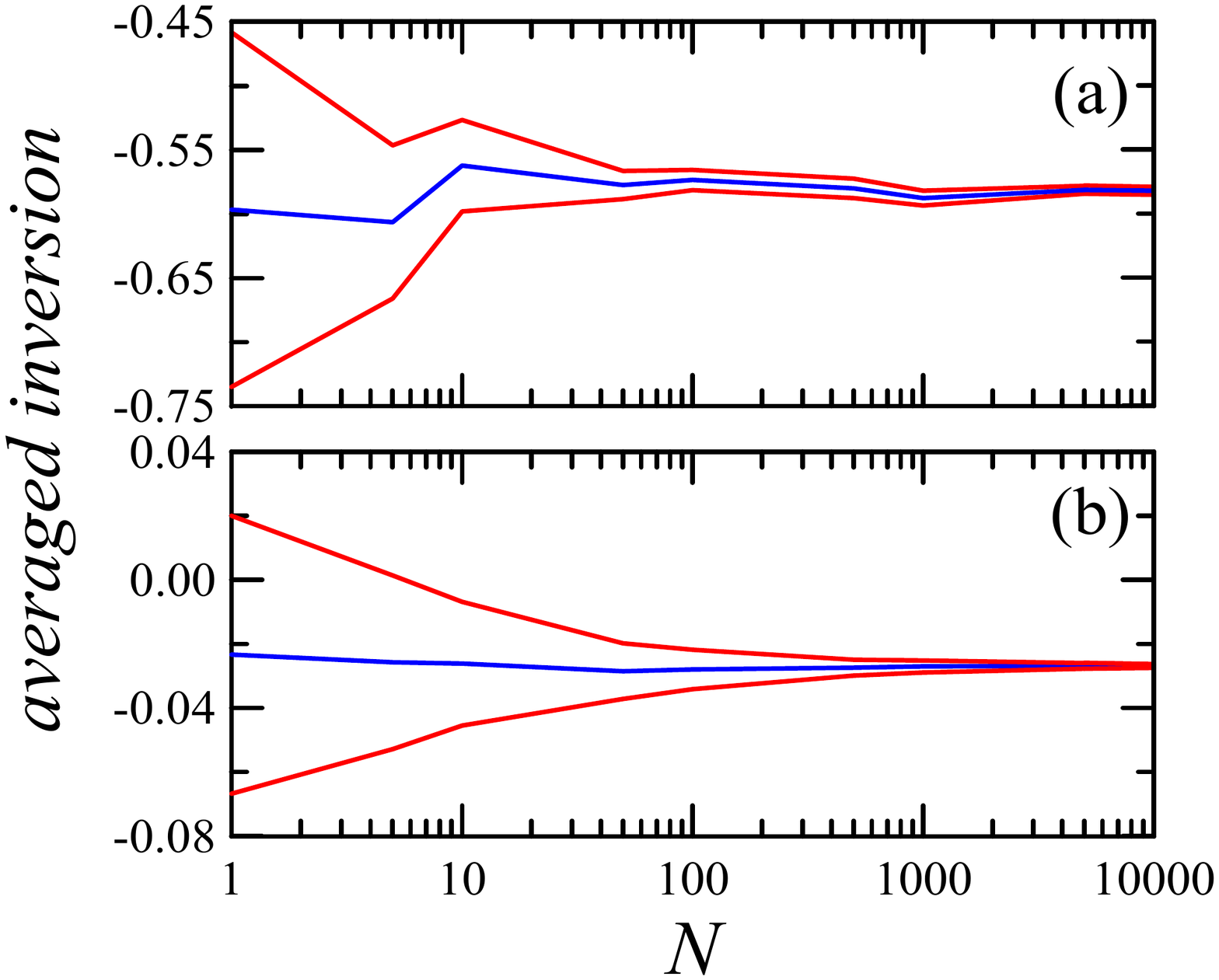}} 
\end{center}
\caption{}
\label{fig:FIG3}
\end{figure}


\begin{thebibliography}{99}
\bibitem{Lorentz} H. A. Lorentz, \textit{The Theory of Electrons} (Teubner,
Leipzig, 1909), Chap. 4.

\bibitem{Einstein} A. Einstein, \textit{Quantum theory of radiation}, Phys.
Z. \textbf{18}, 121-128 (1917); an English translation appears in \textit{\
The World of the Atom}, edited by H. A. Boorse and L. Motz (Basic Books, New
York, 1966), Vol. 2, pp. 888-901.

\bibitem{Milonni} P. W. Milonni and J. H. Eberly, \textit{Lasers} (John
Wiley \& Sons, New York, 1988)

\bibitem{Shore} B. W. Shore, \textit{The Theory of Atomic Coherent
Excitation } (John Wiley \& Sons, New York, 1990).

\bibitem{Loudon} R. Loudon, \textit{The Quantum Theory of Light} (Oxford
University Press, 2000).

\bibitem{Dodd} J. N. Dodd, \textit{Atoms and Light: Interactions} (Plenum
Press, New York, 1991).

\bibitem{nota1} It is important to remark that standard semiclassical theory
does not explain spontaneous emission. This means that the derivation of the 
$A$ coefficient requires quantization of both matter and electromagnetic
field (see, e.g., Ref.\,5). Contrarily, the standard semiclassical theory
allows to derive an expression for the $B$ coefficient, yielding the same
result as obtained with the fully quantized theory, \cite{Loudon}, as we
show here. We must remark, however, that there is at least one formulation
of the semiclassical theory that accounts for spontaneous emission:
Self-field Quantum Electrodynamics, the theory developed by A. O. Barut and
collaborators during the 1980 decade. See A. O. Barut and J. P. Dowling, 
\textit{Self-field quantum electrodynamics: the two--level atom}, Phys. Rev.
A \textbf{41}, 2284-2294 (1990) and references therein.

\bibitem{Barnet} S. M. Barnet and P. M. Radmore, \textit{Methods in
Theoretical Quantum Optics} (Oxford University Press, 2003).

\bibitem{Mandel} L. Mandel and E. Wolf, \textit{Optical Coherence and
Quantum Optics} (Cambridge University Press, 1995).

\bibitem{nota5} Of course it could be also said that for large $\Delta $ it
is the spectral energy density at the Bohr frequency, $W_{21}$, the quantity
governing stimulated processes whilst for large $\Gamma ^{\mathrm{dc}}$ it
is the effective energy density $W_{21}^{\mathrm{eff}}=\zeta W_{21}=\Delta
/\left( 2\Gamma^{\mathrm{dc}}\right) W_{21}$, it is just a question of
taste. Here we choose the $B^{\mathrm{eff}}$ point of view.

\bibitem{Box-Muller} G. E. P. Box and M. E. Muller, \textit{A note on the
generation of random normal deviates,} Ann. Math. Statist. \textbf{29},
610-611 (1958).

\bibitem{Kloeden} P. E. Kloeden and E. Platen, \textit{The numerical
solution of stochastic differential equations} (Springer, 1995).

\bibitem{nota3} For $\Omega\in\mathbb{R}$, it is easy to show that $\mathrm{%
\ Re}\sigma_{21}=0$ in Eqs.\,(\ref{Bloch}). Hence, $\mathrm{Im}\sigma_{21}$
can be identified with $\bar{q}$ in this case.

\bibitem{Fermi} H. Fearn and W. E. Lamb, \textit{Corrections to the golden
rule}, Phys. Rev. A \textbf{43}, 2124-2128 (1991).

\bibitem{nota4} A random process $\xi\left(t\right) $ of zero mean is said
Gaussian if its $n^{\prime}$th-order correlation function verifies  
\begin{gather*}
\left\langle \xi\left( t_{1}\right) \xi\left( t_{2}\right) \ldots \xi\left(
t_{n}\right) \right\rangle = \\
\sum_{\substack{ \mathrm{all}~\left( n-1\right) !!~  \\ \mathrm{pairings} }}
\left\langle \xi\left( t_{1}\right) \xi\left( t_{2}\right) \right\rangle
\left\langle \xi\left( t_{3}\right) \xi\left( t_{4}\right) \right\rangle
\ldots\left\langle \xi\left( t_{n-1}\right) \xi\left( t_{n}\right)
\right\rangle ,
\end{gather*}
for $n$ even, and zero for $n$ odd. i.e., for a Gaussian noise of zero mean
all moments are known if the second order moment is. The Gaussian noise is
said to be white when its second order moment verifies Eq.\,(\ref{corr_xi}),
see Ref.\,9.
\end{thebibliography}
\end{document}